\documentclass[11pt]{article}
\usepackage{moriond,epsfig}

\def\be{\begin{equation}}
\def\bea{\begin{eqnarray}}
\def\eea{\end{eqnarray}}

\newcommand{\jpsi}{J/\psi}
\newcommand{\psip}{\psi^{\prime}}

\newcommand{\pio}{\pi^0}

\newcommand{\pp}{\pi^+\pi^-}
\newcommand{\kk}{K^+K^-}

\newcommand{\gpppp}{\gamma \pi^+\pi^-\pi^+\pi^-}

\newcommand{\gff}{\gamma f_{2} f_{2}}
\newcommand{\gfmfm}{\gamma f_{2}(1270) f_{2}(1270)}

\newcommand{\ff}{f_{2} f_{2}}

\newcommand{\etac}{\eta_{c}}

\newcommand{\ar}{\rightarrow}

\newcommand{\bfg}{\begin{figure}[htpb]}
\newcommand{\efg}{\end{figure}}
\newcommand{\bitm}{\begin{itemize}}
\newcommand{\eitm}{\end{itemize}}
\newcommand{\bnum}{\begin{enumerate}}
\newcommand{\enum}{\end{enumerate}}
\newcommand{\btbl}{\begin{table}[htp]}
\newcommand{\etbl}{\end{table}}
\newcommand{\btbu}{\begin{tabular}[htp]}
\newcommand{\etbu}{\end{tabular}}

\newcommand{\bcl}{\begin{center}}
\newcommand{\ecl}{\end{center}}

\newcommand{\beq}{\begin{equation}}
\newcommand{\eeq}{\end{equation}}
\newcommand{\beqr}{\begin{eqnarray}}
\newcommand{\eeqr}{\end{eqnarray}}

\begin{document}
\vspace*{4cm}
\title{RECENT $\jpsi$ RESULTS FROM BES II}

\author{ N.F. ZHOU \\ (FOR THE BES COLLABORATION)}

\address{Institute of High Energy Physics, Beijing, China, 100049}

\maketitle\abstracts{ Multiquark searches have been a hot topic in
recent years. Some threshold enhancements were observed during the
analysis of 58M $\jpsi$ events collected by BES II detector. These
enhancements can't be explained by the standard model and may be
candidates of multiquark. BES also contribute a lot to the study
of light scalar mesons, such as $\sigma$,$\kappa$,$f_{0}(1370)$
and $f_{0}(1710)$ etc. Study of the excited baryon states and
measurements of some $\jpsi$ decays are also done with the $\jpsi$
events of BES. }

\section{Introduction}
58M $\jpsi$ events have been collected by BES II detector. Some
analysis based on these data have been done and there are some
important results on the multiquark search, light scalar mesons,
excited baryon states and measurements of $\jpsi$ decays.

\subsection{Multiquark candidates at BES}\label{subsec:prod}

According to the naive quark Model, hadrons consist of 2 or 3
quarks. However, QCD allows new forms of hadrons, such as
multi-quark states,hybrids($q\bar{q}g$,qqqg) and
glueballs(gg,ggg).

\begin{figure}[htbp]
  \begin{minipage}[t]{0.5\linewidth}
    \begin{center}
    {\psfig{figure=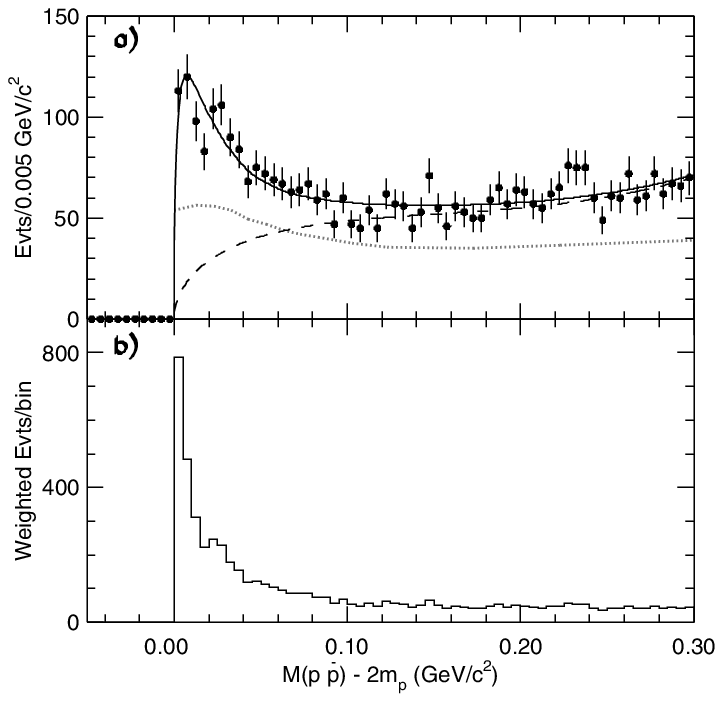,height=2.1in} \caption{The near
threshold $M_{p\bar{p}}-2m_{p}$ distribution for the $\jpsi \ar
\gamma p\bar{p}$ event sample. \label{fig:gpp}}}
    \end{center}
  \end{minipage}
  \begin{minipage}[t]{0.5\linewidth}
    \begin{center}
    {\psfig{figure=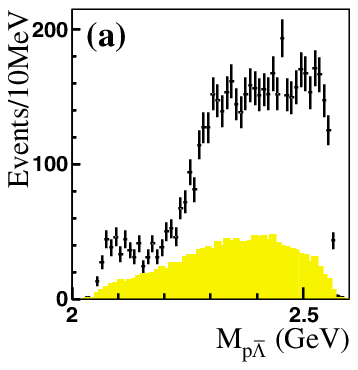,height=2.1in} \caption{The points with
error bars indicate the measured $p\bar{\Lambda}$ mass spectrum
from $\jpsi \ar pK^{-}\bar{\Lambda}+c.c.$ decays. The shaded
histogram indicates phase space MC events(arbitrary
normalization). \label{fig:plamdak}}}
    \end{center}
  \end{minipage}
\end{figure}

We observe a narrow enhancement near $2m_{p}$ in the invariant
mass spectrum of $p\bar{p}$ pairs from radiative $\jpsi\ar \gamma
p\bar{p}$ decays, as shown in Fig. \ref{fig:gpp}.~\cite{gpp} No
similar structure is seen in $\jpsi\ar \pio p\bar{p}$ decays. The
enhancement can be fit with either an S- or P-wave Breit-Wigner
resonance function. In the case of the S-wave fit, the peak mass
is below $2m_{p}$ at
$M=1859^{+3}_{-10}(stat)^{+5}_{-25}(syst)MeV/c^{2}$ and the total
width is $\Gamma <30MeV/c^{2}$ at the $90\%$ confidence level.
These mass and width values are not consistent with the properties
of any known particle.

It is, therefore, of special interest to search for possible
resonant structures in other baryon-antibaryon final states. The
Belle Collaboration observed a near-threshold enhancement in the
$p\bar{\Lambda}$ mass spectrum from $B\ar p\bar{\Lambda}\pi$
decays.~\cite{belle} In the $\jpsi$ data collected by BES II, an
enhancement near the $M_{p}+M_{\Lambda}$ mass threshold is
observed in the combined $p\bar{\Lambda}$ and $\bar{p}\Lambda$
invariant mass spectrum from $\jpsi \ar pK^{-}\bar{\Lambda}+c.c.$
decays, as shown in Fig. \ref{fig:plamdak}.~\cite{plamda} It can
be fit with an S-wave Breit-Wigner resonance with a mass
$m=2075\pm12\pm5$ MeV and a width of $\Gamma=90\pm35\pm9$ MeV. It
can also be fit with a P-wave Breit-Wigner resonance. Evidence for
a similar enhancement is also observed in $\psip \ar
pK^{-}\bar{\Lambda}+c.c.$ decays
 from BES II data.

During the analysis of $p\bar{\Lambda}$ threshold enhancement in
$\jpsi \ar pK^{-}\bar{\Lambda}+c.c.$ decays, we also noticed an
extra-ordinary enhancement near the threshold of the invariant
mass spectrum of $K^{-}\bar{\Lambda}+c.c.$. Partial Wave Analysis
is carried out to understand its properties. Its mass varies
within $1.5\sim1.65$GeV, and its width is about $70\sim 110$MeV.
The most important is that it has a very large branching ratio of
$\sim 2\times10^{-4}$,which indicates an extra-ordinary large
coupling to $K^{-}\bar{\Lambda}$ channel.

\subsection{Light Scalar Mesons}

There have been hot debates on the existence of $\sigma$ and
$\kappa$. Using the 58M $\jpsi$ events from BES II, the decay
$\jpsi \ar \omega \pp$ is studied.~\cite{sigma} At low $\pi\pi$
mass, a large broad peak due to the $\sigma$ is observed.Two
independent analyses are performed, and different
parameterizations of the $\sigma$ pole are applied. The mass and
width of the $\sigma$ are different when using different
parameterizations. However the pole position of $\sigma$ is
stable. Different analysis methods and different parameterizations
of the $\sigma$ amplitude give consistent results for the $\sigma$
pole. From a simple mean of the six analyses, the pole position of
the $\sigma$ is determined to be
$541\pm39-\emph{i}(252\pm42))MeV$.

Based on 58M $\jpsi$ events from BES II, $\bar{K}^{*}(892)K^{
+}\pi^{-}$ channel from $\kk\pp$ data is studied. A clear low mass
enhancement in the invariant mass spectrum of $K^{ +}\pi^{-}$ is
observed. The low mass enhancement does not come from background
events of other $\jpsi$ decay channels, nor from phase space
effect. The scalar resonance $\kappa$ is highly required in the
analysis. Two independent partial wave anlyses have been performed
in $\bar{K}^{*}(892)K^{ +}\pi^{-}$ channel. Both analyses favor
that the low mass enhancement is an isospinor scalar resonant
state. The mass and width are $878\pm60^{+64}_{-55}$ and
$499\pm109^{+55}_{-87}$MeV, respectively in average for results in
both analyses.

Lattice QCD predicts the $0^{++}$ scalar glueball mass at
$1.5\sim1.7$GeV. $f_{0}(1500)$ and $f_{0}(1710)$ are good
candidates. We also studied some other light scalar mesons with
the BES data, such as $f_{0}(980)$.

A partial wave analysis is presented of $\jpsi\ar\phi\pp$ and
$\jpsi\ar\phi\kk$. The $f_{0}(980)$ is observed in both sets of
data, and parameters of the $Flatt\acute{e}$ formula are
determined accurately:
$M=965\pm8(stat)\pm6(syst)MeV/c^{2}$,$g1=165\pm10\pm15MeV/c^{2}$,
$g2/g1=4.21\pm0.25\pm0.21$.

\begin{figure}[htbp]
  \begin{minipage}[t]{0.5\linewidth}
    {\psfig{figure=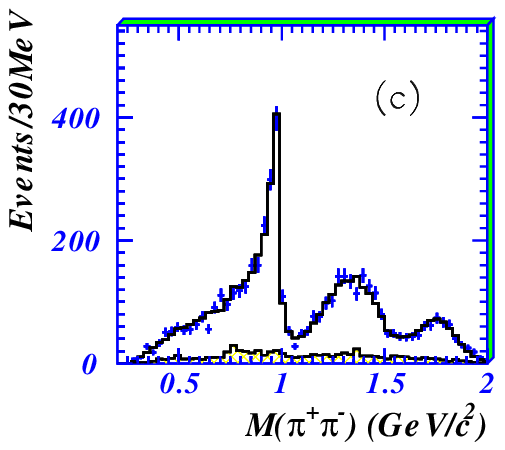,height=2.1in} \caption{The invariant
mass distribution of $\pp$ from $\jpsi\ar\phi\pp$ decays. The
upper histogram shows the maximum likelihood fit and the lower one
shows background. \label{fig:phipipi}}}
  \end{minipage}
  \begin{minipage}[t]{0.5\linewidth}
    {\psfig{figure=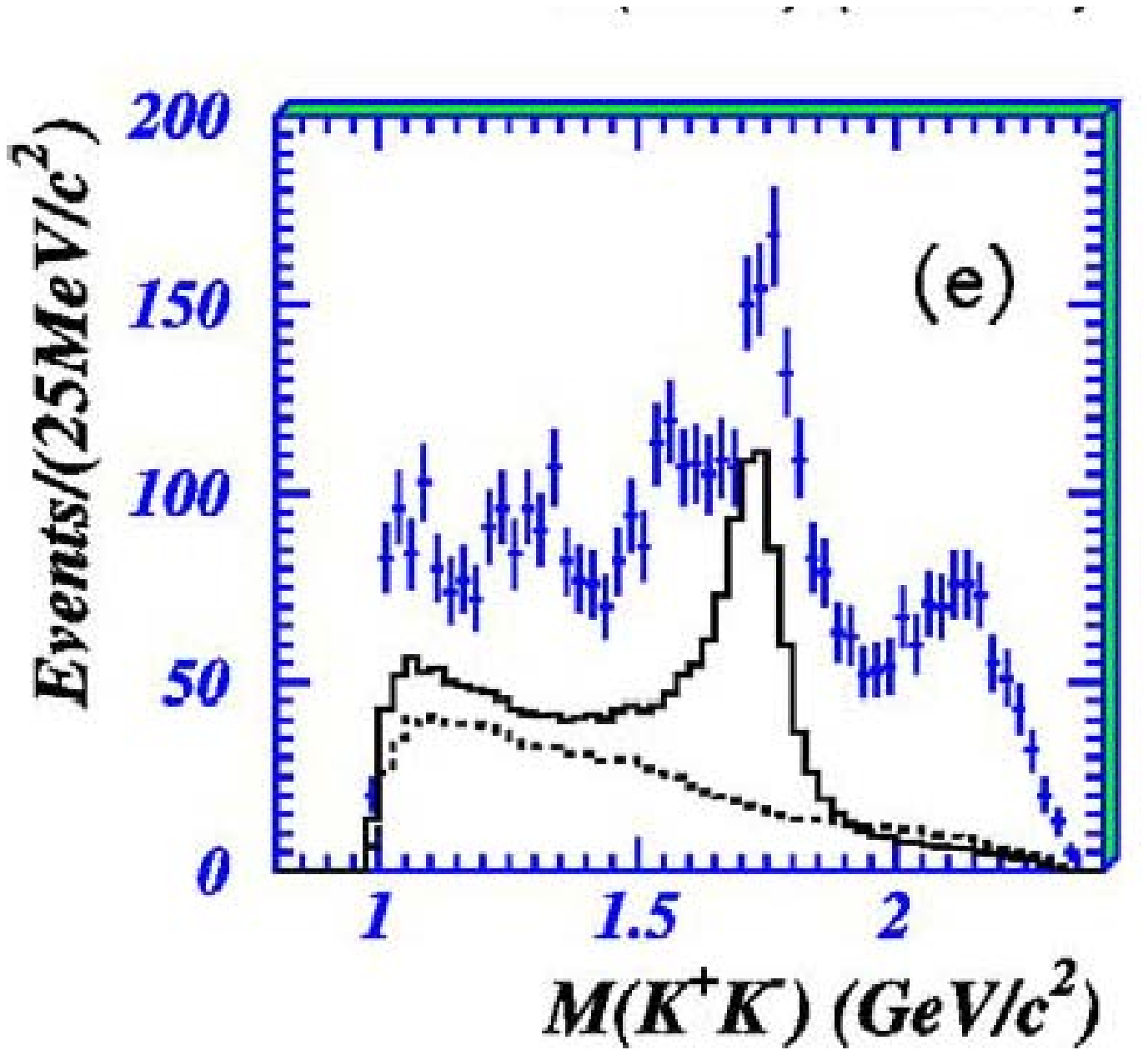,height=2.1in} \caption{The invariant
mass distribution of $\kk$ from $\jpsi\ar\omega\kk$ decays. The
 histogram shows the mass projection of $f_{0}$ and the dashed curve shows
 the $\sigma\ar \kk$ S-wave contribution. \label{fig:omega_kk}}}
  \end{minipage}
\end{figure}

The $\jpsi\ar\phi\pp$ data sample also exhibit a strong $\pi\pi$
peak centered at $M=1335MeV/c^{2}$, as shown in Fig.
\ref{fig:phipipi}. It may be fitted with $f_{2}(1270)$ and a
dominant $0^{+}$ signal made from $f_{0}(1370)$ interfering with a
smaller $f_{0}(1500)$ component. There is evidence that the
$f_{0}(1370)$ is resonant, from interference with $f_{2}(1270)$.
According to the partial wave analysis, the mass of $f_{0}(1370)$
is $M=1350\pm50MeV/c^{2}$ and width is $265\pm40MeV/c^{2}$. In the
analysis of $\jpsi\ar\phi\kk$, the magnitude of the signal due to
$f_{0}(1370)\ar\kk$ in the PWA fit gives a branching fraction
ratio $$\frac{Br(f_{0}(1370)\ar K\bar{K})}{Br(f_{0}(1370)\ar
\pi\pi)} = 0.08\pm0.08 $$ There is no evidence of $f_{0}(1370)$ in
the $\pi\pi$ spectrum of $\jpsi\ar\omega\pp$.

In the analysis of $\jpsi\ar\omega\kk$, there is a conspicuous
signal for $f_{0}(1710)\ar\kk$, as shown in Fig.
\ref{fig:omega_kk}.~\cite{omegakk} A $K\bar{K}$ fit for
$f_{0}(1710)$ with J = 0 yields $M=1738\pm30MeV/c^{2}$ and
$\Gamma=125\pm20MeV/c^{2}$. Earlier BES II data on $\jpsi\ar\gamma
\kk$ and $\jpsi\ar\gamma K^{0}_{s}K^{0}_{s}$ gave
$M=1740\pm4(stat)^{+10}_{-25}(syst)MeV/c^{2}$ and
$\Gamma=166^{+5+15}_{-8-10}MeV/c^{2}$.~\cite{gkk} The branching
ratio for $\jpsi\ar\omega f_{0}(1710),f_{0}(1710)\ar\kk$ is
$(6.6\pm1.3)\times10^{-4}$. From a combined analysis with
$\omega\pp$ data, we find at the 95$\%$ confidence level
$$\frac{Br(f_{0}(1710)\ar \pi\pi)}{Br(f_{0}(1710)\ar
K\bar{K})} < 0.11 $$ In the analysis of $\jpsi\ar\gamma \pi\pi$,
there is a scalar around 1765MeV. It may come from $f_{0}(1790)$,
$f_{0}(1710)$ or a mixture of $f_{0}(1710)$ and $f_{0}(1790)$.

According to the OZI rule, in $\jpsi$ decays, an $\omega$ or
$\phi$ signal determines the $u\bar{u}+d\bar{d}$ or $s\bar{s}$
component, respectively. However, there are some unusual
properties of the $f_{0}(1370)$ and $f_{0}(1710)$. The
$f_{0}(1370)$ dominantly decays to $\pi\pi$(not to KK)$\ar
u\bar{u}+d\bar{d}$, but it is mainly produced together with
$\phi$(not $\omega$). The $f_{0}(1710)$ dominantly decays to KK(
not to $\pi\pi$)$\ar s\bar{s}$, but it is mainly produced together
with $\omega$(not $\phi$).

\subsection{Excited baryon states}\label{subsec:fig}
\begin{figure}
\begin{center}
 \psfig{figure=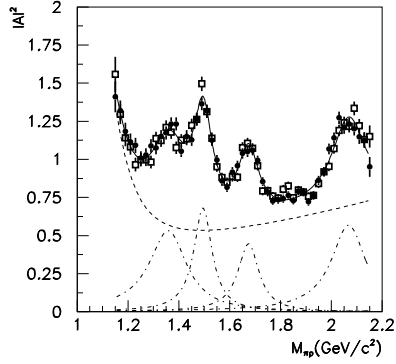,height=2.in} \caption{ Data divided by
Monte Carlo phase space vs $p\pi$ invariant mass for $\jpsi \ar
\bar{p}\pi^{-}\bar{n}$(solid circle) and $\jpsi \ar
\bar{p}\pi^{+}n$(open square), compared with our fit(solid curve).
The contributions of each resonance peak are shown by the
dot-dashed lines in the same figure. The dashed line is the
contribution of background terms. \label{fig:n_star}}
\end{center}
\end{figure}

Using the 58M $\jpsi$ events of BES, more than 100 thousand
$\jpsi\ar p\pi^{-}\bar{n}+c.c.$ events are obtained. Besides two
well known $N^{*}$ peaks at 1500MeV and 1670MeV, there are two
new, clear $N^{*}$ peaks in the $p\pi$ invariant mass spectrum
around 1360 MeV and 2030 MeV, as shown in Fig. \ref{fig:n_star}.
They are the first direct observation of the $N^{*}(1440)$ peak
and a long-sought "missing" $N^{*}$ peaks above 2 GeV in the $\pi
N$ invariant mass spectrum. A simple Breit-Wigner fit gives the
mass and width for the $N^{*}(1440)$ peak as $1358\pm6\pm16MeV$
and $179\pm26\pm50$MeV, and for the new $N^{*}$ peaks above 2 GeV
as $2068\pm3^{+15}_{-40}MeV$ and $165\pm14\pm40$MeV.

\subsection{Measurement of $\jpsi$ decays}\label{sec:plac}

Some branching ratios of $\jpsi$ decays are measured for the first
time in BES. The decay $\jpsi\ar\gfmfm\ar\gpppp$ is observed for
the first time and its branching ratio is measured to be
$Br(\jpsi\ar\gff)=(9.5\pm0.7\pm1.6)\times10^{-4}$. ~\cite{gff} The
branching fraction of $\jpsi\ar\gamma\etac,\etac\ar\ff$ is also
measured to be $Br(\jpsi\ar\gamma\etac)\cdot
Br(\etac\ar\ff)=(1.3\pm0.3^{+0.3}_{-0.4})\times10^{-4}$. The
branching fractions of $\jpsi\ar 2(\pp) \eta$ and $\jpsi\ar 3(\pp)
\eta$ are measured for the first time to be
$(2.26\pm0.08\pm0.27)\times 10^{-4}$ and
$(7.24\pm0.96\pm1.11)\times 10^{-4}$, respectively.

\section*{Acknowledgments}

These works of BES collaborations are done by many people inside
and outside BES collaboration. Thank all people who contributed to
these works.

\end{document}